\documentclass[a4paper,11pt]{report}
\usepackage{amsmath}
\usepackage{epsfig}
\usepackage{graphicx}
\usepackage[round]{natbib}

\title{Applications of Bayesian Probability Theory in Astrophysics}
\author{Brendon James Brewer}
\date{\vspace{1cm}
\includegraphics[scale=0.1]{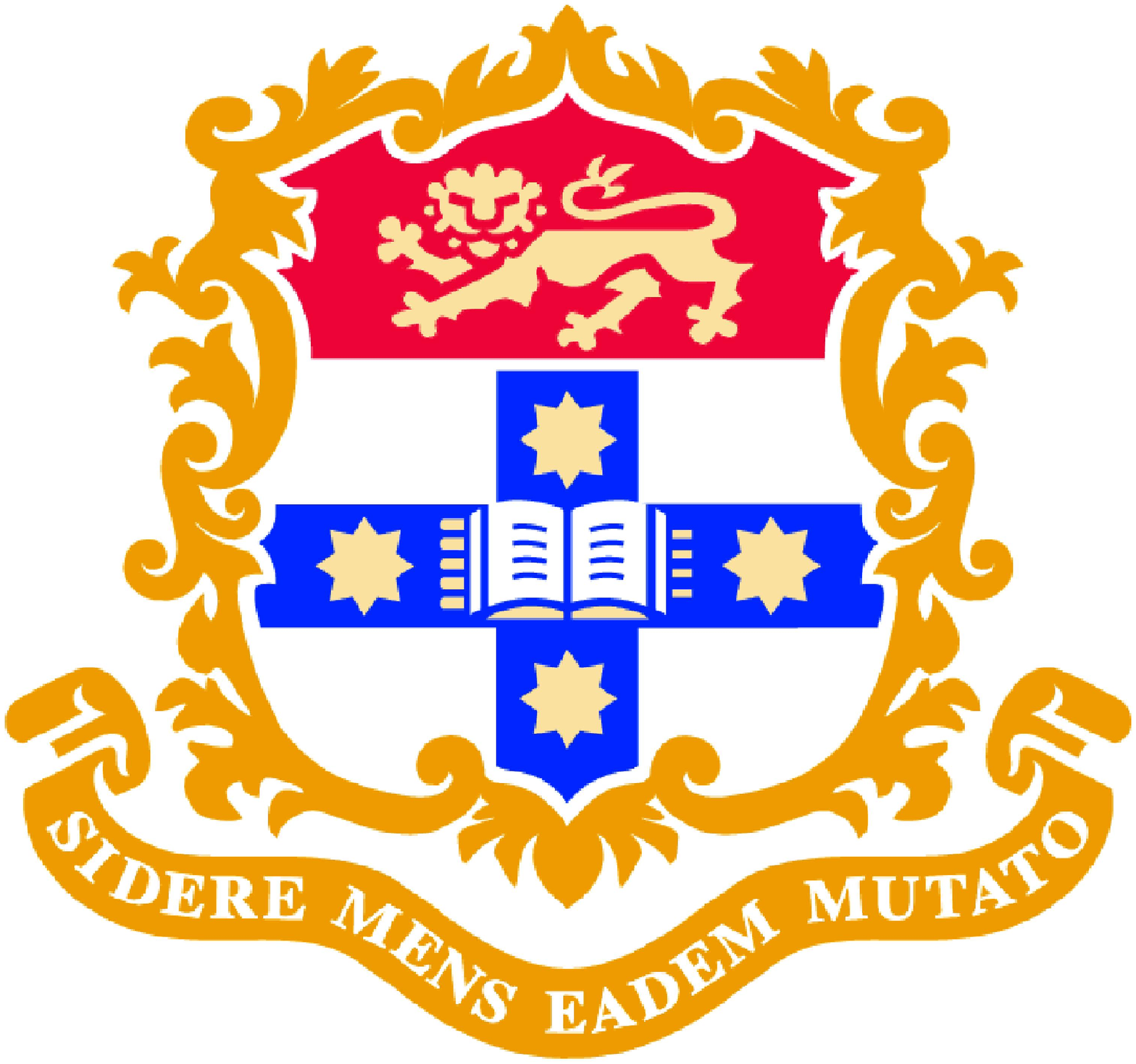} \\\vspace{1cm}
A thesis submitted for the degree of \\Doctor of Philosophy \\at the University of Sydney \\\vspace{1cm}
May, 2008 
}

\begin{document}
\maketitle
\noindent
\begin{abstract}
Bayesian Inference is a powerful approach to data analysis that is based almost entirely on probability theory. In this approach, probabilities model {\it uncertainty} rather than randomness or variability. This thesis is composed of a series of papers that have been published in various astronomical journals during the years 2005-2008. The unifying thread running through the papers is the use of Bayesian Inference to solve underdetermined inverse problems in astrophysics. Firstly, a methodology is developed to solve a question in gravitational lens inversion - using the observed images of gravitational lens systems to reconstruct the undistorted source profile and the mass profile of the lensing galaxy. A similar technique is also applied to the task of inferring the number and frequency of modes of oscillation of a star from the time series observations that are used in the field of asteroseismology. For these complex problems, many of the required calculations cannot be done analytically, and so Markov Chain Monte Carlo algorithms have been used. Finally, probabilistic reasoning is applied to a controversial question in astrobiology: does the fact that life formed quite soon after the Earth constitute evidence that the formation of life is quite probable, given the right macroscopic conditions?
\end{abstract}

\vspace{3cm}
\begin{center}
{\Large\bf{Statement of Originality}}
\end{center}
This thesis describes work carried out in the Institute of Astronomy, within the School of Physics, University of Sydney, between February 2005 and May 2008. This thesis contains no material that has been presented for a degree at this or any other university, and, to the best of my knowledge, contains no copy or paraphrase of work published by another person, except where duly acknowledged in the text.\vspace{0.5cm}\newline

...................................\hspace{3cm}......................... \newline
\hspace{1cm}Brendon J. Brewer\hspace{4.2cm}		Date \newline

\begin{center}
{\Large\bf{Publications Included in this Thesis}}
\end{center}
Several chapters of this thesis have been published in peer-reviewed journals. Others are currently in the review process, and are listed as ``submitted''. All of the first-author papers consist of research conducted by myself, in consultation with my supervisor Geraint Lewis, or other collaborators where applicable. The papers made use of astronomical data that were taken by others but provided to me in order to carry out this work. All of the text and figures of these papers were also written by me, with the exception of minor corrections and the occasional paragraph. For the paper on the radio Einstein Ring PSS2322+1944 (the only non-first author paper included in this thesis), the data were taken by Dominik Riechers and his team, who contacted us to undertake the gravitational lensing analysis. A significant fraction of the work contained in the paper (basically, everything related to lensing and Bayesian inference) was carried out by myself, producing Figures 4, 5 and 7 (and the results shown in Figure 6). The text describing the gravitational lens inversion was written by me, and forms a significant part of the paper. The other parts of the paper were predominantly written by Dominik Riechers.\vspace{1cm}

Strong Gravitational Lens Inversion: A Bayesian Approach. {\it B.~J. Brewer} and G.~F. Lewis, 2006, Astrophysical Journal 637, 608-619. \newline

The Einstein Ring 0047-2808 Revisited: A Bayesian Inversion. {\it B.~J. Brewer} and G.~F. Lewis, 2006, Astrophysical Journal 651, 8-13. \newline

Unlensing HST Observations of the Einstein Ring 1RXS J1131-1231: A Bayesian Analysis. {\it B.~J. Brewer} and G.~F. Lewis, Monthly Notices of the Royal Astronomical Society, in press. \newline

A Molecular Einstein Ring at $z=$4.12: Imaging the Quasar Host Galaxy of PSS2322+1944 Through a Cosmic Lens. D.~A Riechers, F.~Walter, {\it B.~J. Brewer}, C.~L. Carilli, G.~F. Lewis, F.~Bertoldi and P.~Cox, the Astrophysical Journal, in press. \newline

Bayesian Inference from Observations of Solar-Like Oscillations. {\it B.~J. Brewer}, T.~R. Bedding, H. Kjeldsen and D. Stello, 2007, Astrophysical Journal 654, 551-557. \newline

Implications of the Early Formation of Life on Earth. {\it B.~J. Brewer}, submitted to Astrobiology. \newline
\newpage
\begin{center}
{\Large\bf{Publications Not Included in this Thesis}}
\end{center}
The following papers have been published or submitted for publication and are related to the work presented in this thesis. They have not been included because they are either periphirally related to this thesis, or my contribution to the work was relatively minor.\\

Getting Your Eye In: A Bayesian Analysis of Early Dismissals in Cricket\\
{\it B.J. Brewer}, submitted to the Australian and New Zealand Journal of Statistics. ArXiv: 0801.4408\\

Solar-Like Oscillations in the Metal-Poor Subgiant Nu Indi: II. Acoustic Spectrum
and Mode Lifetime\\
F.~Carrier, H.~Kjeldsen, T.~R.~Bedding, {\it B.~J.~Brewer}, R.~P.~Butler, P.~Eggenberger, F.~Grundahl, C.
~McCarthy, A.~Retter and C.~G.~Tinney, 2007, Astronomy and Astrophysics 470, 1059-1063.\\

Solar-Like Oscillations In The G2 Subgiant Beta Hydri From Dual Site
Observations \\
T.~R. Bedding, H.~Kjeldsen, T.~Arentoft, F.~Bouchy, J.~Brandbyge, {\it B.~J.~Brewer}, R.~P.~Butler, J.~Christensen-Dalsgaard, T.~Dall, S.~Frandsen, C.~Karoff, L.~L.~Kiss, M.~J.~P.~F.~G.~Monteiro, F.~P.~Pijpers, T.~C.~Teixeira, C.~G.~ Tinney, I.~K.~Baldry, F.~Carrier and S.~J.~O'Toole, 2007, The Astrophysical Journal 663, 1315-1324. \\

Solar-Like Oscillations in the Metal-Poor Subgiant Nu Indi: Constraining the
Mass and Age using Asteroseismology\\
T.~R.~Bedding, R.~P.~Butler, F.~Carrier, F.~Bouchy, {\it B.~J.~Brewer}, P.~Eggenberger, F.~Grundahl, H.~Kjeldsen,
C.~McCarthy, T.~B.~Nielsen, A.~Retter and C.~G.~Tinney, 2006, The Astrophysical Journal 647, 558-563.

\newpage
\begin{center}
{\Large\bf{Acknowledgements}}
\end{center}
\vspace{1cm}

Firstly, I would like to thank my supervisor, Geraint Lewis, for all of the advice, optimism, guidance and (occasional) debugging. Tim Bedding, who effectively became my associate supervisor for part of the project, is also thanked for his patience and enthusiasm. Officemates past and present are acknowledged in approximate reverse chronological order: Matt Francis, Richard Lane, Madhura Killedar, Juliana Kwan, Daniel Yardley, Chris Hales, Berian James, Luke Barnes, Blair Conn, Tyoma Tuntsov - thanks for all of the company and conversation, some frivolous, some serious, some overly ambitious\footnote{I now realise that numerically solving the Einstein Field Equations is quite hard.}. I'd also like to thank Elizabeth Mahony -just because. I hope the following people will enjoy seeing their name in print as much as I enjoy their friendship: Adrian Daly, Olivia Ross, Renee Bugden, Phillip Crowley, Bill Sherwood, Craig ``Langy'' Lancaster, Alexandra Volosinas, Luke Smith, Meagan Walsh, Daryn Voss (especially for his random cricket SMSs and Spamusement cartoons), Sacha van Albada, Pete Rowney, Victa Phu, Kate Morris (especially for looking after Zazu), John Yasmineh, Anelie Walsh, and everyone who should be on this list but isn't.

To my family: Nanna, Dad, Mum, Keagan, Laura, Caity and Rhys - thanks for your support. I would like to thank Debbie, Paul, Danielle, Andrew and Ben Fender, for putting up with my frequent visits. I have been inspired by the writings of Edwin T. Jaynes, David J. C. MacKay, Richard Dawkins, Douglas Hofstadter and many other authors and scientists. Some of their influence (albeit with only a small fraction of their knowledge) may be detected throughout this thesis. 

I am grateful to the Health Bar in the Wentworth Building for providing the best yoghurt on campus, and Karen for always serving said yoghurt with a friendly smile. I recommend the ``plain'' yoghurt, with muesli added. The following musicians made the writing of this thesis more enjoyable than it otherwise would have been: Elvis Costello, Pat Metheny, Kate Bush, The Arlenes, Brian Wilson, Randy Newman, Cold Chisel, Tori Amos, Trio T\"oyke\"at and many others.

I would also like to thank the people who have collaborated with me on various projects contained in this thesis: I learnt a lot, and hope you did too. During my candidature I have been financially supported by an Australian Postgraduate Award and a Denison Merit Award from the School of Physics. For this I would like to thank the taxpayers of Australia. I have also made extensive use of various infrastructure (roads, electricity, plumbing, etc) that are necessary but rarely acknowledged; additionally I have used a large amount of software written by others, particularly the GNU/Linux operating system and Matlab\footnote{I also used Mozilla Firefox far too much.}.

\begin{quote}
I could carry on for hours and I am sure I have missed many people, but I have to get this page to the binders.

- Geraint F. Lewis, {\it Gravitational Lensing}, PhD Thesis
\end{quote}

\tableofcontents

\chapter{Introduction}\label{intro}
\vspace{-1cm}
\begin{quote}
I refuse to answer that question on the grounds that I don't know the answer.

    - Douglas Adams
\end{quote}
\section{Forward and Inverse Problems}
Science proceeds by comparing the predictions of theories with observations or experiments designed to test them \citep{jeffreys}. If a theory correctly predicts the outcome of some observations or an experiment, our level of confidence in the theory is increased, otherwise the theory becomes less plausible\footnote{Often, the theory that is being penalised is not some important or fundamental theory, but the hypothesis that the experiment is doing what it was thought to be doing. In less words, the experiment could be wrong.}. In recent years, the increase in computing power has allowed theorists to carry out increasingly realistic simulations of physical phenomena, predicting very specific details that are often beyond the reach of current observational and experimental limits. For instance, in the field of n-body simulations, it is now possible to answer the question ``given certain initial conditions at some time $t_1$, and some assumed laws of physics, what will be the state of the system at a later time $t_2$?'' \citep{1988csup.book.....H}. However, it is much harder to answer the reverse question: given all of the observed data, what can now be said about the laws of physics that have been operating?

This kind of scenario is often referred to as an {\it inverse problem} \citep{inverse}. It is often possible to solve a forward problem, reasoning from some physical assumptions to a prediction for what would be observed. But reasoning from an observation back to the correct model is difficult, and usually underdetermined in the sense that there are many possible explanations for a given data set. A typical observation rules out many theories but is consistent with many others. While many specific inverse problems may be solvable using techniques invented separately for each problem, there is a general framework for solving all problems of this type. This framework is introduced in the following section.

\section{What is Bayesian Inference?}
When probability theory is taught, it is usually introduced without a great deal of discussion as to the meaning or interpretation of the quantity ``probability''. Most introductory applications are based around games of chance - coin flipping, dice rolling, and card games. In these cases, some intuitively obvious
assumptions are made, such as the assumption that each of the six sides of a die will appear with probability $1/6$ and each toss is independent. The probabilities of more complex events can then be calculated by applying the mathematical rules of probability theory. These rules will now be stated. For any two propositions or events $A$ and $B$, the {\it product rule} states
\begin{equation}\label{productRule}
P(A,B) = P(A)P(B|A)
\end{equation}
where $P(A,B)$ is the probability that \textbf{both} $A$ and $B$ are true, and $P(B|A)$ is the probability that $B$ is true, given that $A$ is true. The {\it sum rule} relates $P(A)$ to $P(\tilde{A})$, the probability that $A$ is false.
\begin{equation}
P(A) + P(\tilde{A}) = 1
\end{equation}
As usual, all probabilities are real numbers in the interval $[0,1]$.

The mathematical content of probability theory consists of the above rules and their consequences. However, for the scientist or applied mathematician, the question of the interpretation of probability is of considerable importance, because it determines the range of applicability of the above equations.

\subsection{Subjective Probabilities}
It is not the purpose of this thesis to repeat the long history of the controversies over the meaning of probability; the interested reader is referred to \citet{jaynes} and references therein. Although Jaynes was a partisan, his career took place in the mid to late 20th century, at a time when the Bayesian view was much less mainstream than it is today.

An important derivation of the rules of probability was provided by \citet{cox2,cox}. He was seeking a generalisation of standard Boolean logic to take uncertainty into account. Applying a few basic consistency criteria implies that, if the {\it degree of plausibility} of a hypothesis $H$ is modelled by a real number, then the consistency criteria are only met if the rules for combining {\it plausibilities} are equivalent to rules of probability theory - in other words, the laws of probability are the unique rules for reasoning in the presence of uncertainty\footnote{The rules of probability can also be justified from requirements on rational betting behaviour, with no reference to repititions, frequencies or averages \citep{2002PhRvA..65b2305C}.}. Thus, probability theory can be used as a mathematical model of our state of knowledge about the plausibility of various hypotheses. Any hypothesis with a probability of $1$ is certain to be true, those with probability $0$ are certain to be false, and any number between these describes some intermediate level of certainty.

Of course, the degree of plausibility of any proposition depends on the information that is taken into account and the assumptions that are being made. Hence, Bayesian probabilities are always conditional probabilities. In general, the probability of a hypothesis $H$ given information or assumptions $I$, denoted $P(H|I)$, is different to the probability of $H$ given different information $J$, denoted $P(H|J)$. It is certainly the case that different people can disagree about the probability of the same hypothesis - but only if they have different information, or are making different starting assumptions. It is in this sense that probabilities are subjective - but ideally they should be assigned based on logical reasoning and all of the available evidence. Occasionally, the nature of the evidence can be complicated and qualitative, and assumptions like ``that seems implausible, so we will assign probability 1/100'' will have to be made at the beginning of the calculation. If the outcome depends on any ad-hoc judgments such as these, then the answer to the problem is that it depends on what you assume, which is a situation not unique to Bayesian probability theory.

Consider two propositions $H$ and $D$. Expanding the joint probability of $H$ and $D$ in the two ways allowed by Equation~\ref{productRule} gives
\begin{equation}
P(H,D|I) = P(H|I)P(D|H,I) = P(D|I)P(H|D,I)
\end{equation}
where everything is conditional on the background information and/or assumptions $I$. Whilst $I$ can be omitted from equations to increase readability, it is always present, and neglecting this can lead to confusion when comparing two calculations that appear equivalent, but implicitly depend on different assumptions. Rearranging, we obtain {\it Bayes' Theorem}
\begin{equation}\label{bayes}
P(H|D,I) = \frac{P(H|I)P(D|H,I)}{P(D|I)}
\end{equation}
If $H$ is a hypothesis about nature, and $D$ is a statement about some observed data, this equation describes how to calculate our plausibility for $H$ in the light of the extra data $D$ --- it is equal to the probability of $H$ before taking into account the data, times $P(D|H,I)$, a measure of how well $H$ would have predicted $D$ to occur, and divided by $P(D|I)$, the chance that this data would be observed whether the hypothesis was true or false.

\section{Using Bayes' Theorem for Estimation}
It is straightforward to generalise Equation~\ref{bayes} to the case where there are a large number of competing mutually exclusive hypotheses. In practise, the most common situation is that we are interested in learning the value of some quantity or set of quantities $\theta$. Then the hypotheses we wish to test (and calculate the probabilities of, given some data) are of the form \{${\theta = 0.300},{\theta = 0.301},{\theta = 0.302},...$\} (for discrete variables; continuous variables are described by density functions where the probability of the true value being in any finite interval is the integral of the density over that interval). In this case, a version of Bayes' theorem for probability distributions can be used.

To estimate unknown quantities (denoted collectively by $\theta$) from measurements of other quantities $D$, we model our state of knowledge by a joint distribution over the space of possible values for $\theta$ and $D$. Then we condition on the values of $D$ that were actually observed, to calculate the {\it posterior distribution} for $\theta$:
\begin{equation}\label{densityBayes}
p(\theta|D,I) \propto p(\theta|I)p(D|\theta,I)
\end{equation}
The $p$'s in this equation can be either discrete probability distributions or continuous probability density functions, depending on the possible values that the $\theta$ and $D$ variables can take. Usually, rather than choosing a model distribution for $p(\theta|D,I)$ directly, it is much easier to assign the two distributions on the right side of Equation~\ref{densityBayes}; $p(\theta|I)$ is called the {\it prior distribution} and describes our state of knowledge about $\theta$ before taking into account the current data, and $p(D|\theta,I)$, which models our predictions about the data given the parameters of interest. When the actual data $D$ are known and fixed, $p(D|\theta,I)$ becomes a function of $\theta$ only and is called the {\it likelihood function}.

\begin{figure}[!h]
\begin{center}
\includegraphics[scale=0.8]{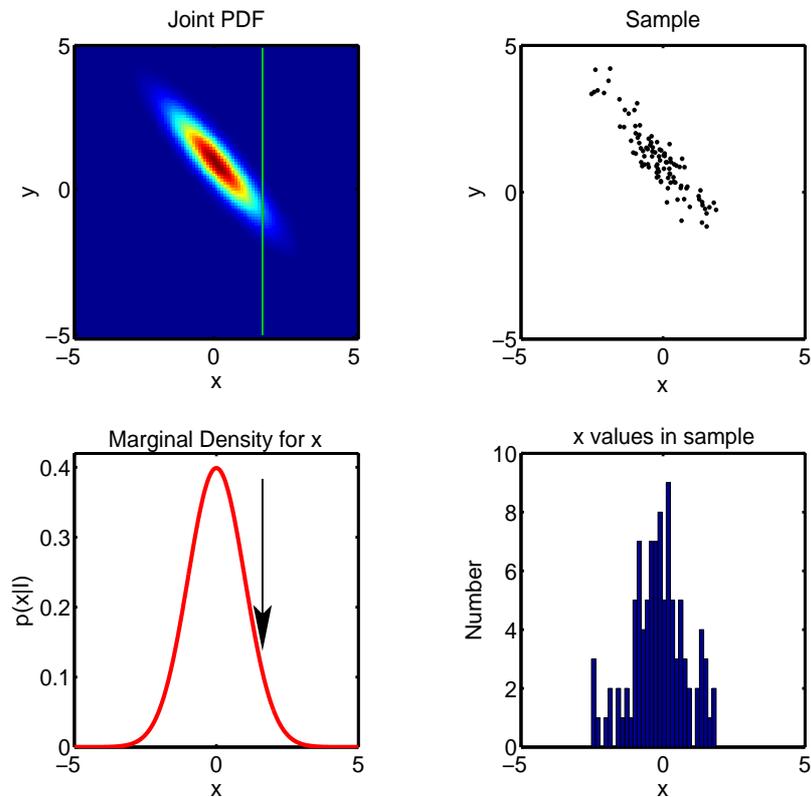}
\caption{The top left panel shows a joint Gaussian probability density function $p(x,y|I)$ for two variables, $x$ and $y$, and the corresponding marginal density for $x$ is shown below. If this was the posterior distribution for two parameters, a sample from this distribution, such as that shown in the right hand panels, would suffice to estimate quantities such as marginal estimates and error bars for $x$. The sample size in this case was 100, and increasing the sample size would make the sample a accurate approximation to the continuous densities in the left hand panels.\label{marginalisation}}
\end{center}
\end{figure}

If $\theta$ is made up of some parameters of interest, $x$, and some uninteresting parameters, $y$, a posterior probability distribution for $x$ alone can be calculated by the process of {\it marginalisation}. Integrating a joint probability distribution over one of the variables yields the {\it marginal distribution} for the other variables:
\begin{equation}\label{marg}
p(x|D,I) \propto \int p(x,y|D,I) dy
\end{equation}
where the integration is over all possible values of $y$; Figure~\ref{marginalisation} is a graphical representation of the process of marginalisation. The top left panel shows a joint probability density function $p(x,y|I)$ for two variables, $x$ and $y$. The result of integrating the joint density along the green line $x = 2$ becomes the value of $x$'s marginal density $p(x|I)$ at $x=2$. From the right hand panels, it is apparent that if we can acquire a random sample of points in the $(x,y)$ plane sampled from the joint distribution (here, $N=100$), simply ignoring the $y$ values results in a sample from the marginal distribution for $x$; this is the motivation behind Markov Chain Monte Carlo algorithms, which are introduced in section~\ref{mcmc}.
For more details on the basic philosophy and practice of Bayesian Inference, the introductory textbooks by \citet{sivia} and \citet{gregory} are recommended. For readers already familiar with the basics, \citet{ohagan} is a useful reference. \citet{jaynes} provides entertaining and thought provoking discussions of fundamental principles, and \citet{mackay} is full of interesting examples. Applications to cosmology are reviewed by \citet{trotta}.

\section{Markov Chain Monte Carlo}\label{mcmc}
The main principles of Bayesian Inference are simple and have been described in the previous section. The inputs are the choices of realistic models for the prior distribution and the sampling distribution/likelihood function, and the output of interest is a distribution that is proportional to their product. However, there are often significant challenges involved in carrying out the necessary calculations, particularly the calculation of marginal distributions using Equation~\ref{marg}. In complicated problems, the parameter space (set of possible $\theta$ values) is often high dimensional - in common applications it can range from a few variables to thousands. Integrating functions of many variables is difficult, and in this case the integrand is often sharply peaked, if the data constrain the parameters very well.

To address this challenge, Markov Chain Monte Carlo (MCMC) techniques have been developed and are becoming very popular \citep{gilks}. The basic idea is to use random number generation in order to sample points from a distribution that is equal to the posterior distribution. If each point in the parameter space can be thought of as a model, this means we use a computer to randomly generate a sample of models for the data, according to their probability distribution given the data and prior information. Then, properties of the sample can be used to assess the level of uncertainty remaining about $\theta$, simply by considering measures of the diversity of the $\theta$'s that were returned by the algorithm.

If $\theta$ is multidimensional, such a sample can easily be used for marginalisation\footnote{Although MCMC and other posterior sampling methods may seem slow at times, they are effectively calculating integrals over very high dimensional spaces - a consoling thought.} - the marginal distribution for a parameter can be viewed simply by plotting a histogram of the variable of interest, ignoring the values of the others.

The mathematical definition of a Markov Chain is a set of random variables $\{X_1,X_2,X_3,...\}$ with the special property that the joint probability distribution
\begin{equation}
p(X_1,X_2,...|I) = p(X_1|I)p(X_2|X_1,I)p(X_3|X_2,X_1,I)...
\end{equation}
is such that the conditional distribution for $X_{i+1}$ depends only on the value of $X_i$ and not on the previous history of the chain. Then the joint distribution can be written as
\begin{equation}
p(X_1,X_2,...|I) = p(X_1|I)p(X_2|X_1,I)p(X_3|X_2,I)p(X_4|X_3,I)...
\end{equation}
The idea behind MCMC is to generate an instance of a Markov Chain, starting at some point $X_1$ in the parameter space, and generating subsequent states from the probability distribution $p(X_{i+1}|X_i,I)$, called the transition kernel \citep{neal}. The transition kernel is specially constructed such that as $i$ increases, the marginal probability distributions for all of the $\{X_i\}$ tends towards the distribution that we wish to sample (usually the posterior distribution for some parameters given data). Less formally, {\it an MCMC algorithm explores the parameter space by a random walk, but spends more time in regions of higher probability, such that in the long run, the amount of time spent in any region is proportional to the amount of probability in that region}. Due to the random walk aspect of MCMC, the generated sample of points is not independent, but for the purposes of calculating parameter estimates and error bars, this is not a major drawback. As long as the chain is simulated for a long enough time to have generated a few tens of essentially independent samples, this is usually good enough.
There are many different ways of constructing chains that have this property \citep{neal}. A sufficient, but not necessary, condition is that the chain is ergodic (meaning that any state can eventually be reached from any other), and satisfies the {\it detailed balance} condition:
\begin{equation}
\frac{p(X_{i+1} = y| X_i = x)}{p(X_{i+1} = x| X_i = y)} = \frac{f(y)}{f(x)}\label{detailed}
\end{equation}
where $f$ is proportional to the probability density we would like the chain to explore.
\subsection{The Metropolis-Hastings Algorithm}
The Metropolis-Hastings method is one of the simplest and most popular MCMC algorithms \citep{1953JChPh..21.1087M, hastings}. Suppose that the prior density and likelihood as functions of $\theta$ are $\pi(\theta)$ and $L(\theta)$ respectively, and these can be evaluated for any point $\theta$ in the parameter space. If the current state of the simulation is $\theta_i$, the next state, $\theta_{i+1}$ is chosen as follows: A new value $\theta'$ is generated, chosen from a proposal distribution $q(\theta'|\theta)$. Then the next state of the chain, $\theta_{i+1}$ is equal to $\theta'$ with probability $\alpha$ (called the acceptance probability); otherwise, with probability $1-\alpha$, the chain remains in the same state, so $\theta_{i+1}=\theta_i$. The acceptance probability $\alpha$ is given by
\begin{equation}
\alpha = \mbox{min}\left[1,\frac{q(\theta|\theta')}{q(\theta'|\theta)} \times \frac{\pi(\theta')L(\theta')}{\pi(\theta)L(\theta)}\right] \label{metropolis}
\end{equation}
In words, this means that the current state is randomly perturbed (according to $q(\theta'|\theta)$), and accepted if the new state of the chain is at a region of higher posterior density $\pi L$. If the posterior density is lower at the proposed point, it still has a chance of being accepted, that chance being equal to the ratio of the new density to the old. The extra factor involving $q$ is there to ensure detailed balance if the proposal density is asymmetric. Commonly, it is symmetric, meaning that the chance of proposing $\theta'$ if the current state is $\theta$ is the same as proposing the reverse move if the current state were $\theta'$. It is straightforward to show that the Metropolis updating rule satisfies detailed balance (Equation~\ref{detailed}), and is therefore a valid MCMC algorithm. This means that long Markov chains simulated from the acceptance rule in Equation~\ref{metropolis} will explore the parameter space with the fraction of time spent in any volume being proportional to the total amount of probability contained in that volume.

In practice, we need to choose a proposal distribution. A common method is to add a small, normally distributed perturbation to the current state. If $\theta$ is one dimensional, this would mean that $q(\theta'|\theta) \sim N(\theta,\sigma^2)$. The typical size of the perturbation, $\sigma$, is chosen by the user, and affects the performance of the method. If $\sigma$ is too small, most of the proposed moves will be accepted, but the exploration of the parameter space will be via a slow random walk. If $\sigma$ is too large, most of the proposed moves will be rejected, and the chain will stay in the same state for a long time. A convenient but somewhat wasteful approach is to randomise the width of the proposal to a new value at every step.  Generally, it is a good idea to aim for about 50\% of the proposed moves to be accepted. In multidimensional parameter spaces, proposal changes can be made that only perturb the value of one parameter, or that perturb multiple parameters simultaneously.

Of course, if the starting point for the chain, $X_1$, is far from the bulk of the probability distribution, it may take a long time before samples are effectively being taken from that distribution. This depends on the nature of the target probability distribution and the Metropolis-Hastings proposal distribution chosen. Generally, if the distribution is not multimodal, an arbitrary starting point in a low probability region will eventually wander uphill due to the selective pressure\footnote{In this respect, MCMC simulations resemble evolution by natural selection \citep{dawkins}, and can also be used to construct optimisation algorithms, such as simulated annealing \citep{1983Sci...220..671K}} exerted by the acceptance probability of Equation~\ref{metropolis}. This initial uphill climb of an MCMC simulation is referred to as the burn-in period, and $\theta$ values sampled during this period are usually discarded. Detecting whether an MCMC chain has converged to the target distribution is a difficult problem and there are no guarantees \citep{gilks}. Visual inspection of the output is often good enough, and in the author's opinion, is more useful than any formal criterion.

\begin{figure}[!h]
\begin{center}
\includegraphics[scale=0.55]{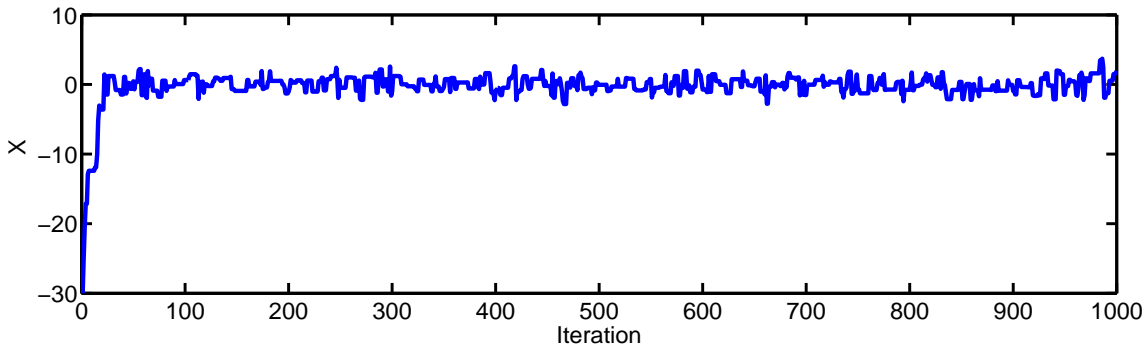}
\caption{An example of an MCMC run applied to sampling from a standard normal distribution. The burn-in period is short in this case, about 100 steps, because the target density is unimodal and the local gradient always influences the chain to move towards the mode. This helpful property is common but not universal.\label{metropolispic}}
\end{center}
\end{figure}

An example of the Metropolis-Hastings algorithm is presented in Figure~\ref{metropolispic}. The target distribution which the sample should come from was chosen to be a standard normal distribution (mean 0, standard deviation 1). After an initial burn-in period of about 100 iterations, the chain successfully samples from the target distribution. Although adjacent samples are not independent, this simulation contains about as much information about the target density as 100 independent samples would. Amusingly, the presence of rejections in the Metropolis-Hastings algorithm can cause the output to resemble a city skyline; this is not the origin of the term {\it Metropolis algorithm}\footnote{Apparently, Geraint Lewis thought this was the case for some time.}. If necessary, MCMC algorithms that are more efficient than straightforward Metropolis-Hastings are available \citep{murray}.

Throughout this thesis, most of the MCMC uses the Metropolis-Hastings algorithm as described above, and an extension called {\it Reversible Jump} MCMC \citep{ohagan, green}. Reversible jump is used when there is uncertainty, not just about the value of the parameters, but also about the number of parameters that should be in the model. For example, in Chapter 7, there is uncertainty not just in the frequencies of some stellar oscillation modes, but also how many modes there are in the first place. This is achieved by including proposal transitions that add an extra component to the model or remove a component from the model (i.e. increase or decrease the dimensionality of the parameter vector, respectively). The adding and removing proposals are chosen in order to satisfy detailed balance with respect to the prior distribution, and are then accepted or rejected using the likelihood ratio only. For example, when adding an extra component to the model, the extra component is generated from its prior distribution (given the other components). When removing a component, a particular component is chosen at random. This is how the MCMC methods in Chapters 4-7 were implemented.

The primary application of MCMC methods considered in this thesis is to an inverse problem in the study of gravitational lenses. When a massive object lies along the line of sight from the observer to a distant source in the universe, the gravitational field of the massive object bends the light rays coming from the distant source. As a result, the observed image does not accurately reflect the actual morphology of the background source. In addition, the image that we see is degraded by blurring caused by atmospheric effects (for example, scintillation, where random fluctuations in the conditions in the Earth's atmosphere cause background point sources to ``twinkle'', and extended sources to blur when viewed over timescales greater than fractions of a second) and instrumental effects such as the diffraction of the received light as it enters the telescope. Additionally, the telescope electronics and the fact that we can sometimes observe only small numbers of photons are some common causes of additional random noise in the image. The challenge is to take a blurred and noisy observed image and to simultaneously infer the original undistorted source profile and the mass profile of the intervening ``gravitational lens''. To do so will require some basic gravitational lensing theory, which is introduced in the next section.

\chapter{Gravitational Lensing}
\begin{quote}
It's...bending. And I'm...watching it bend.

	- Gerry McCambridge, in {\it Psychokinetic Silverware}, Gerry and Banachek
\end{quote}
This chapter presents a brief summary of basic gravitational lensing theory. It is not comprehensive, but covers the main ideas that are used in the subsequent papers. For introductory purposes, the review article by \citet{wambsganss} is recommended. Alternatively, the reference work by \citet{schneider} is denser in historical content and technical details.

\section{Basic Lensing Theory}
\begin{figure}
\begin{center}
\includegraphics[scale=0.9]{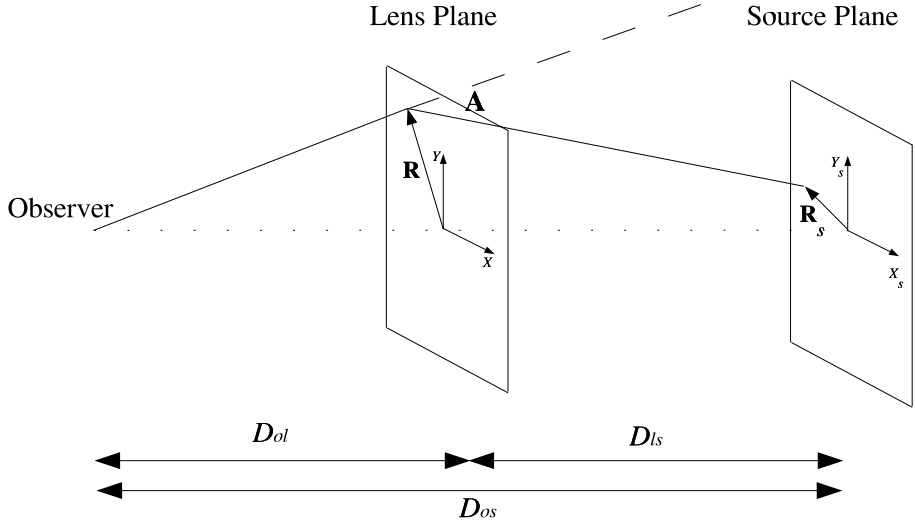}
\caption{A standard gravitational lensing situation. The lensing object is assumed to have its mass distributed in the {\it lens plane}, and the light rays undergo a sharp deflection at this plane. This approximation is valid as long as $D_{ol}$ and $D_{ls}$ are much greater than the extent of the lens mass distribution along the line of sight \citep{schneider}. \label{geometry}}
\end{center}
\end{figure}

The existence of the phenomenon of gravitational lensing is one of the most important predictions of Einstein's theory of general relativity, implying that the paths of light rays are deflected as the light passes near a massive object \citep{schneider}. This implies that if a distant point source is observed, and a massive object lies between the source and the observer, the apparent position of the source is changed. A diagram illustrating this idea is shown in Figure~\ref{geometry}. The distances $D_{ol}$, $D_{os}$ and $D_{ls}$ are the angular diameter distances \citep{hogg} from the observer to the lens, the observer to the source, and the lens to the source. Note that, in general, $D_{os} \neq D_{ol} + D_{ls}$ on cosmological scales due to general relativistic effects.

If a light source is positioned at a point $\mathbf{R}_s \equiv (X_s,Y_s)$ in the {\it source plane} and a light ray from this source is observed, it will appear to have arrived from a different place, due to the deflection of the ray by the gravitational lens. The apparent angular position of the light source in the sky is changed, but its new position is described by its apparent coordinates $\mathbf{R} \equiv (X,Y)$ in the {\it lens plane}. Using simple geometry applied to Figure~\ref{geometry}, it can be shown that the relationship between $\mathbf{R}$ and $\mathbf{R}_s$ is given by:
\begin{equation}
\mathbf{R}_s = \frac{D_{os}}{D_{ol}}\mathbf{R} - D_{ls}\mathbf{A(R)}\label{lenseqn}
\end{equation}
where $D_{os}$, $D_{ol}$ and $D_{ls}$ are the angular diameter distances from the observer to the source, observer to the lens, and lens to source respectively. The deflection angle $\mathbf{A(R)}$ is a vector function defined over the lens plane, and its form depends on how the mass in the gravitational lens is distributed over the source plane. Note that equation~\ref{lenseqn} gives a unique source plane position $\mathbf{R}_s$ for a given image plane position $\mathbf{R}$. In general, though, a unique inverse function does not exist, so any particular position $\mathbf{R}_s$ in the source plane can be mapped to multiple positions in the lens plane. This is the mathematical reason why the background source is often multiply imaged in gravitational lens systems.

It can be shown from the general theory of relativity that the deflection angle $\mathbf{A_0(R)}$ for a point mass lens at the origin of the lens plane is given by an inverse $1/R$ law:
\begin{equation}
\mathbf{A_0(R)} = \frac{4GM}{c^2|\mathbf{R}|}\mathbf{\hat{R}}\label{greens}
\end{equation}
where $G$ is Newton's gravitational constant, $c$ is the speed of light in vacuum and $M$ is the mass of the point gravitational lens. Interestingly, a similar result can be derived from either special relativity or Newtonian mechanics [assuming a $1/r^2$ gravitational field and that light is made up of particles that travel at $c$, see \citet{schneider}], although it gives a result that is smaller by a factor of $2$, in conflict with observations.

The result for a point mass lens is generalised to arbitrary continuous mass distributions $\rho(\mathbf{R})$ by integrating the deflection angle due to each small mass element of the gravitational lens, using Equation~\ref{greens} as a Green's function:
\begin{equation}
\mathbf{A(R)} = \frac{4G}{c^2}\int_{L}\frac{\mathbf{R - R'}}{|\mathbf{R - R'}|^2}\rho(\mathbf{R}')d^2\mathbf{R}'\label{contrho}
\end{equation}
For a point mass lens located at $\mathbf{R} = (0,0)$ and a point source at the origin of the source plane, [$\mathbf{R}_s = (0,0)$], the observed image would be a circular ring with an angular radius equal to the {\it angular Einstein Radius}
\begin{equation}
\theta_0 = \sqrt{\frac{4GM}{c^2}\frac{D_{ls}}{D_{ol}D_{os}}}\label{er}
\end{equation}
This is because any point in the lens plane at a distance $D_{ol}\theta_0$ (an Einstein Radius in the lens plane) from the origin is mapped via~\ref{lenseqn} onto the origin of the source plane. The Einstein Radius provides a convenient length scale for discussing gravitational lensing. Typically, the coordinates $\mathbf{R}$ and $\mathbf{R}_s$ are replaced by the scaled coordinates:
\begin{equation}
    \mathbf{r}  \equiv  \frac{\mathbf{R}}{D_{ol}\theta_0}
\end{equation}
\begin{equation}
    \mathbf{r}_s  \equiv  \frac{\mathbf{R}_s}{D_{os}\theta_0}
\end{equation}
The scaled deflection angle $\mathbf{a}$ is defined as:
\begin{equation}
	\mathbf{a} = \frac{D_{ls}}{D_{os}\theta_0}\mathbf{A}
\end{equation}
With these changes, the lensing equation~\ref{lenseqn} takes following simple form, referred to as the {\it normalised lens equation}:
\begin{equation}
	\mathbf{r}_s = \mathbf{r} - \mathbf{a(r)}\label{scaledlens}
\end{equation}
In addition, the surface mass density profile $\rho(\mathbf{R})$ is replaced by the dimensionless surface mass density $\sigma(\mathbf{r})$ measured in units of $\frac{M}{(D_{ol}\theta_0)^2}$, where $M$ is the mass used in the definition of the Angular Einstein Radius. This change eliminates the prefactors from the scaled version of Equation~\ref{contrho}, which now takes the simple form\footnote{Many authors choose to define the dimensionless density by dividing the actual density by the {\it critical density}, which is the density where a uniform disc of matter maps all points in the lens plane to the same point in the source plane. The critical density is $\sigma_{crit} = \pi^{-1}$.}:
\begin{equation}
\mathbf{a(r)} = \int_{L}\frac{\mathbf{r - r'}}{|\mathbf{r - r'}|^2}\sigma(\mathbf{r}')d^2\mathbf{r}'
\end{equation}
Since the deflection angle field $\mathbf{a}(\mathbf{r})$ is built up of $\mathbf{\hat{r}}/|\mathbf{r}|$ kernels, it must have zero curl and can therefore be written as the gradient of a potential $\phi(x,y)$. In terms of the lensing potential, the lens equation is:
\begin{equation}
\mathbf{r}_s = \mathbf{r} - \nabla \phi
\end{equation}

\subsection{Computation of Extended Images}
Most of the gravitational lensing studies in this thesis are concerned with extended images. In these cases, the source is modelled by a non-negative surface brightness function $S(x_s,y_s)$ over the source plane. Surface brightness conservation implies that the observed image due to lensing is:
\begin{equation}\label{surface}
I(x,y) = S(x_s(x,y),y_s(x,y)) = S(x - \alpha_x(x,y),y - \alpha_y(x,y))
\end{equation}
If the goal is to predict how such an image would appear on the sky, it would need to be convolved by a point spread function (PSF) and then divided into pixels, each pixel taking an amount of light equal to the integral of the surface brightness function $I(x,y)$ over the pixel.

If the source is coincidentally aligned exactly behind a lens galaxy whose projected mass distribution is circularly symmetric, the symmetry results in an image of a complete ring, called an Einstein Ring. Slight deviations from this exact scenario result in almost circular, almost complete rings, as seen in Figure~\ref{ring}. More complex lens mass distributions, such as galaxy clusters, can produce complex image configurations (Figure~\ref{cluster}).

\subsection{Magnification}
Equation~\ref{surface} is a statement of the conservation of surface brightness. If the flux per unit angular area on the sky is not changed by an intervening gravitational lens, how can gravitational lensing produce magnifications? The answer is that gravitational lenses can produce a change in the apparent area of a source on the sky. Consider a very small image located at $(x,y)$ in the image plane, having area $dx~dy$. Its position in the source plane is given by equation~\ref{lenseqn}, and its area is $dx_sdy_s$, where the ratio of areas is given by the Jacobian determinant of the mapping defined by equation~\ref{lenseqn}:
\begin{equation}
\frac{dx_s}{dx}\times\frac{dy_s}{dy} = \textrm{det}
\left[\begin{array}{ccc}
1-\frac{\partial\alpha_x}{\partial x} & \frac{\partial\alpha_x}{\partial y}\\
\frac{\partial\alpha_y}{\partial x} & 1-\frac{\partial\alpha_y}{\partial y}\end{array}\right]
\end{equation}
An image may be flipped by lensing, resulting in a negative value for the above ratio of differentials. Also, the magnification is the ratio of the image plane area to the source plane area. Hence, the actual (positive) magnification of an image at $(x,y)$ in the image plane is given by:
\begin{equation}
\mu(x,y) = \left|
\textrm{det}\left[ \begin{array}{ccc}
1-\frac{\partial\alpha_x}{\partial x} & \frac{\partial\alpha_x}{\partial y}\\
\frac{\partial\alpha_y}{\partial x} & 1-\frac{\partial\alpha_y}{\partial y}\end{array} \right] \right|^{-1}
\end{equation}
If a single source is multiply imaged, the total magnification (that is, observed total flux over all images divided by intrinsic flux) is the sum of the $\mu$ values at all of the images.

\subsection{Applications}
One application of gravitational lenses is to use the fact that the source is magnified to our advantage: the gravitational lens acts as a natural telescope. To achieve this, the inverse problem must be solved: that is, we must be able to answer the question ``what source profiles and lens mass profiles could possibly have produced this image?''. One of the primary goals of this thesis has been to develop general techniques for answering this question for any given Einstein Ring image. This was first performed by \citet{1989MNRAS.238...43K} on the radio ring MG1131+0456, and in the intervening years much attention has focused on how to improve the algorithms used in order to extract as much information as possible from the data \citep[e.g.][]{1994ApJ...426...60W, 1996ApJ...465...64W, 2003ApJ...590..673W, 2005PASA...22..128B, 2006MNRAS.372.1187W}. This is the focus of the gravitational lensing related papers in this thesis.

\begin{figure}[!h]
\begin{center}
\includegraphics[scale=0.5]{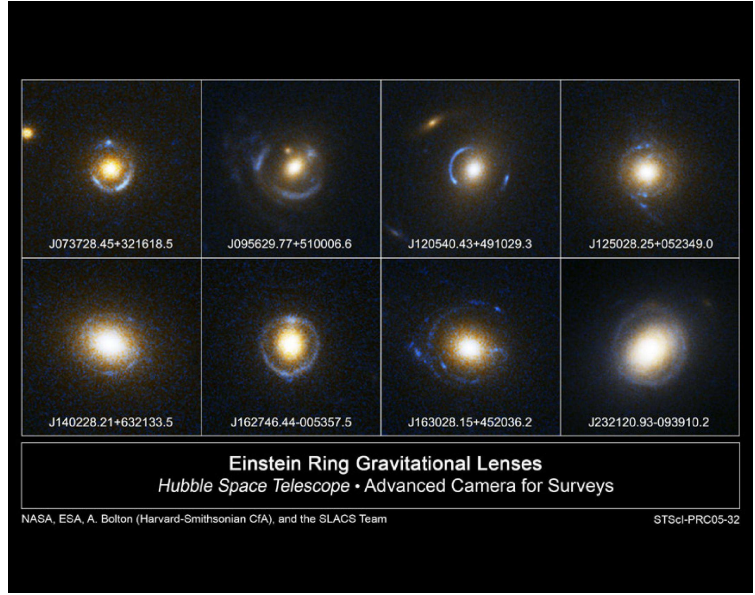}
\caption{Several real images of Einstein Rings (the blue structures) observed with the Advanced Camera for Surveys (ACS) aboard the Hubble Space Telescope. Developing methods for discovering the unlensed source profile is the primary topic of this thesis. The lensing galaxies are also clearly seen in these images. As these are mostly elliptical galaxies they have a much redder appearance.\label{ring}}
\end{center}
\end{figure}

\begin{figure}[!h]
\begin{center}
\includegraphics[scale=0.7]{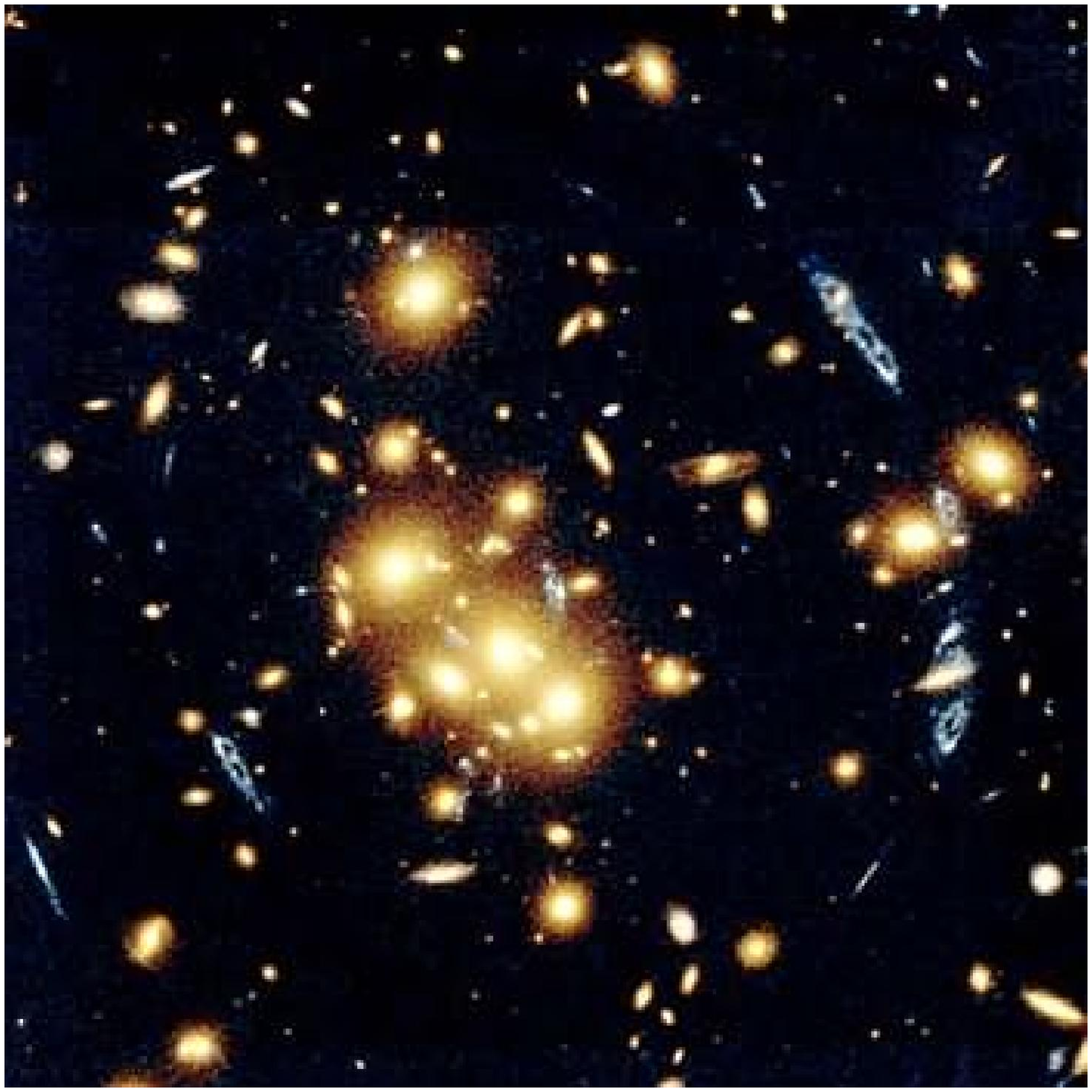}
\caption{Strong gravitational lensing by the galaxy cluster 0024+1654, imaged by the Wide Field Planetary Camera 2 aboard the Hubble Space Telescope \citep{2003ApJ...598..804K}. The multiple images of background galaxies can be found amongst the cluster members; they can be identified by their higher redshift or from a suspiciously lensed shape. This system has been studied thoroughly, with many papers employing different techniques to recover the mass profile of the cluster \citep[e.g.][]{1995ApJ...441...58W,1998ApJ...498L.107T,2000ApJ...542L...1S}, and the source profile \citep{1996ApJ...461L..83C}. \label{cluster}}
\end{center}
\end{figure}

Considerable research has gone into solving a number of similar problems of gravitational lens inversion that focus on the main goal of reconstructing complex lens mass distributions given the positions of either the lensed images or statistical information about the shearing of many background galaxies (the ``weak lensing'' regime). These studies often focus on reconstructing the mass distribution of clusters of galaxies from their lensing effect (Figure~\ref{cluster}).

This lens mass distribution, and particularly the level of substructure present, are important probes of galaxy formation and the nature of dark matter \citep{2005MNRAS.363.1136K,2007ApJ...667..859D}. Recently, much of the effort has gone into so-called ``nonparametric'' modelling, where the unknown lens mass profile is described by a large number of unknown parameters such as pixel values. A popular algorithm for achieving these mass reconstructions is PixeLens \citep{2000AJ....119..439W}, which has seen most of its use in multiply-imaged QSO systems \citep[e.g.][]{2007arXiv0708.2151F}. ``Maximum Entropy'' reconstruction of pixellated mass profiles from weak lensing data has also been proposed and used successfully \citep{1998MNRAS.299..895B, 2002MNRAS.335.1037M, 2003MNRAS.346..489M}. It should be noted that, in this context, ``Maximum Entropy'' has little relation to the principle of maximum entropy as discussed by \citet{jaynes}. Rather, it refers to a particular choice of prior distribution for image reconstruction, that was once thought to have a special theoretical status, but this is no longer thought to be the case \citep{1998mebm.conf....1S}. Nevertheless, reconstructions based on this prior are often impressive.

An example of a modern Bayesian cluster reconstruction algorithm is described by \citet{2007NJPh....9..447J}. While parametric models (such as analytical elliptical mass distributions) may be criticised for assuming too much prior knowledge, pixellated models may be overcorrecting this defect by giving the model too much freedom. Thus, intermediate approaches have also been used, where the mass map is built up from basis functions that are broader and smoother than pixels \citep{2003MNRAS.346..489M, 2007ApJ...655..109P, lensperfect}. For computational reasons, most of these methods do not try to model the entire image and all of the unknown properties of each extended source. Typically, they reduce the data to a set of image positions (and other quantities), and try to find a lens mass distribution that can reproduce the reduced data set.

While we note the existence and success of these methods, they should be considered distinct from the lens reconstruction tasks discussed in this thesis, which focus more on inferring the morphology of the background source. The primary application is galaxy-galaxy lenses (such as those displayed in Figure~\ref{ring}), where it is often assumed that the lens is simple enough to be described by a parametric model.

\chapter{Strong Gravitational Lens Inversion: A Bayesian Approach}
\vspace{-1cm}
{\it B.~J. Brewer and G.~F. Lewis, 2006, Astrophysical Journal 637, 608-619.} \newline
\vspace{-0.5cm}
\begin{quote}
A mathematician may say anything he pleases, but a physicist must be at least partially sane.

    - J. Willard Gibbs
\end{quote}
The first paper in this thesis, {\it Strong Gravitational Lens Inversion: A Bayesian Approach}, introduces an approach based on MCMC for reconstructing extended sources in gravitational lens systems. Previous methods in the literature have usually been based on pixellating the source plane and finding the values of those pixels that minimise the difference (in a least squares sense) between the predicted image and that that has been actually observed \citep{2003ApJ...590..673W}. Since the prediction of an image from a model involves a blurring step, information is lost and the solution to the inverse problem is not unique. This can be solved by using ``regularisation'', where the merit function for minimisation is modified so as to penalise certain kinds of implausible solutions \citep{2006MNRAS.372.1187W}. A simpler approach that is effective if the image is not very high resolution is to model the source parametrically \citep[e.g.][]{2007ApJ...671.1196M}.

In this paper, previous methods based on ``regularisation'' are reinterpreted in a Bayesian context, allowing us to show that the prior information that is implicitly being assumed is not really all that appropriate for astronomical sources - because they did not allow for the fact that we expect many of the source pixels to be dark. The reason for this is that regularisation consists of adding an extra term to the merit function that is to be optimised, and this results in a merit function that looks like the logarithm of the posterior probability density. However, if the regularisation term is interpreted as the logarithm of a prior probability density over the space of pixellated sources, it usually has undesirable features; for example, the linear regularisation commonly used does not specify that negative values for the pixels are not allowed, or even that positive values are more probable than negative ones. In this paper, a basic prior distribution over pixellated sources that takes sky darkness into account is described and applied to simulated data. It should be noted that this paper addresses only a subset of the kinds of problems that might be called ``strong gravitational lens inversion'' - it does not address the reconstruction of complex mass distributions.

All of the code for the numerical work in this paper, and the manuscript itself, was written by myself (BJB), in consultation with my supervisor Geraint Lewis.

\newpage
\chapter{The Einstein Ring 0047-2808 Revisited: A Bayesian Inversion}
\vspace{-1cm}
{\it B.~J. Brewer and G.~F. Lewis, 2006, Astrophysical Journal 651, 8-13.} \newline
\begin{quote}
\vspace{-0.5cm}
How long did it take them to build the telescope?

    - David J. C. MacKay
\end{quote}
This second paper is an application of ideas presented in the first paper (Chapter 3) to real data. The Hubble Space Telescope observations of ER 0047-2808 had already been extensively analysed beforehand, so the results could be directly compared to those obtained by other means. The choice of the prior distribution over pixellated sources was also modified slightly to improve the efficiency of the MCMC exploration. The resulting source reconstruction had a significant improvement in resolution (by about 50 per cent) compared to previous studies, and the lens model parameters were more uncertain because some assumptions could be relaxed and the number of unknown source pixels was increased. Particularly, the central position of the lens model was allowed to be a free parameter - although it was found to be consistent with the central position of the lens galaxy's light profile after all. This paper provides a demonstration of the fact that careful consideration of the available prior information can often lead to improved results in data modelling.

Once again, all of the code for the numerical work in this paper, and the manuscript itself, was written by me (BJB), in consultation with my supervisor Geraint Lewis.

\chapter{Unlensing HST Observations of the Einstein Ring 1RXS J1131-1231: A Bayesian Analysis}
\vspace{-1cm}
{\it B.~J. Brewer, G.~F. Lewis, Monthly Notices of the Royal Astronomical Society, in press} \newline
\vspace{-0.5cm}
\begin{quote}
{\it Hofstadter's Law: }It always takes longer than you expect, even when you take into account Hofstadter's Law.

    - Douglas Hofstadter
\end{quote}
A second optical Einstein Ring observed with the HST is the topic of this third paper. This system posed a number of technical challenges to the technique that was used on 0047-2808. The relatively large number of pixels in the image (121$\times$121), complex source structure and presence of light from the lensing galaxy and the AGN of the source galaxy were all issues that made the inference more difficult than expected. The one source reconstruction that had been done in the literature before this paper appeared used an unusual reconstruction technique \citep{claeskens}. This was done by inferring the lens model parameters from the QSO image positions only, and using this result to back ray-trace the observed image to the source plane. When different parts of the image disagreed about the value of a source pixel, the median value was used.

The claim made by \citet{claeskens} that the QSO images alone were stronger constraints on the lens model than the extended images seemed unlikely given the number and detail of the extended structures; hence, a significant fraction of this paper is devoted to exploring this issue. The conclusion of this comparison was that the QSO images are only strong constraints on the lens parameters if the central position of the lens mass distribution is fixed in advance. Otherwise, the extended images become far more informative. In general, if the extended source is simple (for example, one diffuse component), then extended images do not confer any additional information about the lens than point images. However, if the extended source contains a lot of substructure (as is the case with J1131), the extended images provide vital extra information.

All of the code for the numerical work in this paper, and the manuscript itself, was written by myself (BJB), in consultation with my supervisor Geraint Lewis.

\newpage

\clearpage
\chapter{A Molecular Einstein Ring at $z=$4.12: Imaging the Quasar Host Galaxy of PSS2322+1944 Through a Cosmic Lens}
\vspace{-1cm}
{\it D.~A Riechers, F. Walter, B.~J. Brewer, C.~L. Carilli, G.~F. Lewis, F. Bertoldi, P. Cox, Astrophysical Journal, in press} \newline
\vspace{-0.5cm}
\begin{quote}
OK, one last time. These are {\it small}. But the ones out there are {\it far away}. {\it Small... far away...} ah forget it!

- Father Ted
\end{quote}
The fourth paper in this thesis is a study of a radio Einstein Ring \citep{carilli} at redshift 4.12, seen in the wavelength of the Carbon Monoxide 2--1 transition. The data consisted of seven images at slightly different frequencies, and the image of the ring changes with frequency due to velocity structure of the emitting CO gas cloud. Thus, there is not one single unknown source, but seven, all lensed by a common mass profile. CO is of particular interest because it is able to exist in the same conditions (temperature, etc) as molecular hydrogen - the raw fuel for star formation and the powering of active galactic nuclei.

A modelling challenge encountered in this study was the fact that the noise in the images is correlated on length scales of several pixels. Correlated noise models can be computationally expensive, so the following simplified approach was used: simply increase the size of the ``error bars'' on the data so that only a fraction of the information in the data is being taken into account; where the correct fraction is determined by the effective number of pixels with independent noise values. This approach produced satisfactory results because the large PSF meant that the modelling could not overfit structures smaller than the noise correlation scale.

The results revealed for the first time a detailed multi-wavelength reconstruction of the extended molecular source, which is a thin disk-like structure with a velocity gradient: i.e. one end of the source is redshifted relative to the other end, with an intermediate velocity for the central parts. The optical quasar source does not reside at the centre of this disk, so the CO gas is not orbiting the central black hole of the galaxy in the simplest possible way. This indicates that the structure of the source galaxy is complex, as expected for high redshift star forming galaxies, in standard galaxy formation scenarios.

The observations and data reduction in this paper were undertaken by our collaborators, and I (BJB) wrote and used the lens modelling code, producing Figures 4, 5 and 7 (and the results shown in Figure 6), with input from my supervisor, Geraint Lewis. I also wrote the parts of the text describing the lens inversion technique (section 4) and am responsible for the resultant interpretation of the reconstructed source.

\newpage

\chapter{Bayesian Inference from Observations of Solar-Like Oscillations}
\vspace{-1cm}
{\it B.~J. Brewer, T.~R. Bedding, H. Kjeldsen, D. Stello, Astrophysical Journal 654, 551-557.} \newline
\begin{quote}
On the subject of stars, all investigations which are not ultimately reducible to simple visual observations are...necessarily denied to us...We shall never be able by any means to study their chemical composition.

- August Comte, French philosopher (1835)
\end{quote}
\vspace{0.5cm}
The following paper is one of two in this thesis that is not concerned with the topic of gravitational lens inversion. Asteroseismology is the study of the intrinsic oscillations of stars. In recent years, there has been an explosion in the study of {\it solar-like oscillations} in main sequence stars - this is where the star oscillates in many modes simultaneously and with small amplitude. For example, the sun oscillates in many frequencies with a typical period of $\sim$ 5 minutes \citep{asteroreview}. Knowing the frequencies of the oscillation modes of a star is a powerful constraint on the internal properties of the star. 

This process is analogous to hearing the sound of an unknown musical instrument and using the properties of that sound to infer something about the physical structure of the instrument \citep{kac}. While the sound doesn't provide information about absolutely every aspect of the instrument, it does provide some information. Similarly, stellar oscillations can tell us something, but not everything, about the internal structure of the star. Hence, it is also an underconstrained inverse problem. However, this paper does not deal with the problem of inferring stellar structure from frequencies, it is instead concerned with inference of the frequencies from observational data of the variations of the star's brightness or surface velocity with time - basically, fitting multiple sine waves to noisy data. This is essentially the classic problem of data analysis: given some noisy data, find out whether there is a signal present, and if so, what its properties are.

While conventional Fourier techniques are usually adequate (taking the Fourier transform of something close to sinusoidal yields peaks at the frequencies of the oscillations), they can only be derived from Bayes' theorem under certain assumptions that do not always apply \citep{bretthorst}, and therefore may not be using all of the information in the data. The approach we presented in this paper is more generally applicable, and has been applied to observations of three stars \citep{2006ApJ...647..558B,2007ApJ...663.1315B,2007A&A...470.1059C}. These papers have not been included in this thesis due to space constraints; in any case, my contribution tended to be small and to confirm the results of the conventional analysis. Recently, the method presented in this paper has been criticised on the grounds that we did not take damping and excitation into account in our modelling \citep{2007arXiv0711.0435A}. However, previous approaches based on the power spectrum are also implicitly modelling the signal as sinusoids \citep{bretthorst}, and are also vulnerable to similar criticisms.

The ideas and numerical work presented in this paper were done by myself (BJB), with Tim Bedding and Hans Kjeldsen providing guidance on asteroseismology and particularly on the kinds of assumptions that such a program should and should not make. Dennis Stello produced the simulations of stochastically excited and damped oscillations that were used in the paper.

\chapter{Implications of the Early Formation of Life on Earth}
\vspace{-1cm}
{\it B.~J. Brewer, submitted to Astrobiology.} \newline
\vspace{-0.5cm}
\begin{quote}
It is as true in probability theory as in carpentry that introduction of more powerful tools brings with it the obligation to exercise a higher level of understanding and judgment in using them. If you gave a carpenter a fancy new power tool, he {\it may} use it to turn out more precise work in greater quantity; or he may just cut off his thumb with it. It depends on the carpenter.

- Edwin T. Jaynes
\end{quote}

This paper stemmed from a discussion with Dr Charlie Lineweaver that took place at the Harley Wood Winter School held in June 2004 at Coolangatta, Queensland. Dr Lineweaver had given a talk on the topic of his work in computing the ``Galactic Habitable Zone'', the times and places in the galaxy where conditions such as metallicities and low supernova frequencies are ideal for life \citep{2004Sci...303...59L}. During the talk, it was suggested that we should expect life (at least basic forms) to be common in the galaxy, because it formed on Earth a surprisingly short time after the surface had cooled sufficiently \citep{2002AsBio...2..293L}. Whilst this certainly constituted some evidence in that direction, the confidence with which Dr Lineweaver expressed his conclusion seemed unjustified. After the talk, I mentioned this to Dr Lineweaver; however, my thoughts were not sufficiently well-formed to make a convincing case at the time. This paper presents a more well-developed case, showing that because our own existence is more probable if life forms early, the fact that life {\it did} form early is not conclusive - although it does constitute {\it some} useful evidence in this difficult and often speculative field. In this paper, I revisit the Lineweaver and Davis model and find that their overconfident conclusion resulted from their unintentional use of a very informative prior distribution.

\newpage
\chapter{Conclusions and Further Work}
This thesis has presented an array of applications of Bayesian statistics to several complex data analysis problems in astrophysics. While data analysis methods exist for these problems, in the past these have tended to be based on simple ideas with limited applicability (such as least-squares fitting). The first problem considered was that of inferring the unknown surface brightness profile of a gravitationally lensed galaxy, where the image has also been convolved with a point-spread function and subject to additive noise. Compared to methods based on least squares fitting (with or without regularisation), our method consistently produces sharper (higher resolution) reconstructions of the source; this has been demonstrated strongly in the case of ER 0047-2808. Application of our method to RXS J1131-1231 also produced some different conclusions to previous work, especially regarding the constraining power of the extended images compared to pointlike images. In this system, the large amount of substructure in the extended source is extremely valuable for constraining the lens mass distribution, besides being of interest in its own right. For the radio (carbon monoxide) Einstein Ring PSS2322+1944 at redshift of 4.12, the spectacular multi-wavelength source reconstruction presented in this thesis is the first inversion using the recent new observations by \citet{2007AAS...211.4502R}. The lens inversion for this system is a necessary step towards studying the complex dynamics of the molecular gas in the source galaxy. The number of high quality images of Einstein Rings is increasing rapidly \citep[e.g.][]{slacs} and the approach demonstrated in this thesis should prove valuable for studying them.

The primary reason for the success of this new method is that the number of source pixels to be inferred is large compared to the number of pixels in the observed image. In this regime, the prior information that the method is implicitly assuming becomes important, and ordinary least-squares is implicitly assuming (in a sense) no prior information, not even positivity. Some regularisation formulas, which can be viewed as priors, amount to strange assumptions of prior information (for example, many linear regularisers/Gaussian priors assume each pixel is independent and is just as likely to be positive as it is to be negative). In contrast, this work has placed a greater emphasis on the realism of the prior distribution, rather than its analytical properties. Whilst this comes at the price of making the resulting models more complicated, computing power and clever numerical techniques (in this case, MCMC) have stepped in to make the computations feasible after all.

Of course, all of this source reconstruction work relies on the assumption that simple analytical lens models are applicable. For isolated galaxy lenses, this is probably not a problem, but it would be highly inappropriate to apply this method as it stands to source reconstructions of galaxies lensed by galaxy clusters. Substructure in galaxy cluster lenses is a major topic of interest, and in principle it should be possible to detect substructure with an approach like the one used here. The only reason it cannot be done presently is the immense computing power required (other methods such as PixeLens \citep{2000AJ....119..439W} and LensPerfect \citep{lensperfect} are feasible because they condition on a reduced data set, such as several image positions, rather than conditioning on the value of every pixel in a large image). A multitude of new Bayesian computation techniques are being developed, such as likelihood-free computation \citep[e.g.][]{sisson} that may be able to contribute to progress in this field.

A similar story holds for the astroseismology parts of this thesis: when commonplace data analysis methods are interpreted in Bayesian terms, the conditions that must hold for them to be applicable become more apparent. For example, it has always been recognised that Fourier analysis of a (non-equally-spaced) time series containing a sinusoidal signal may be difficult due to the phenomenon of aliasing. The method presented in this thesis can effectively accomplish automatic de-aliasing and tell us the uncertainty of any conclusion, allowing for the recovery of modes that may have been discarded as noise by a Fourier or CLEAN analysis. However, this method is also explicitly using an assumption of sinusoidal signals \citep{2007arXiv0711.0435A}. Work has commenced on generalising this approach to take into account the ``semi-regular'' behaviour of solar-like oscillation signals (Brewer and Stello 2008, in preparation).

The algorithms used in this thesis tend to be computationally expensive if the number of pixels/light-atoms/modes is large. There are many more advanced MCMC techniques that could be implemented to improve the efficiency of the sampling and thereby extend the range of systems that could be treated using this approach. Two techniques that may be helpful in this regard are Hamiltonian Monte Carlo \citep{neal} and more advanced implementations of Reversible Jump MCMC \citep{green}, including split/merge operations. For gravitational lens inversion, it should be possible to extend the algorithms in order to model the lens in a ``nonparametric'' way, similar to that employed for the source light profile. This would allow for the detection of substructure without making strong prior assumptions.

Finally, the Bayesian formalism was applied to the study of the early formation of life on earth, in order to quantify the significance of this information for the question of whether life is common in the universe. This study highlighted an unintentional assumption in a previous analysis that caused the authors to reach overconfident conclusions. The work presented in this thesis corrects this mistake and shows that the current data do not decisively rule out the possibility that we are alone, although they do disfavour this possibility.

In the limit when all of our data become very good, all of these efforts will become irrelevant: but the most active parts of science are always those where the questions we are asking are {\it not} clearly and unambiguously determined by our current data. For this reason, the question of how to analyse noisy and incomplete data, to combine information from different sources, and to honestly express the implications of that information, will always be with us. As more and more advanced instruments are made, and observations planned, more sophisticated data analysis tools and a deeper understanding of their rationale will be required. The recent rise in Bayesian activity amongst astronomers suggests that these challenges are widely recognised. Consequently, Bayesian Inference continues to become an important and widely-used tool of modern astronomical research.

\bibliographystyle{plainnat}

\appendix

\end{document}